\documentstyle[12pt]{article}
\title{Erratum\\ Quantum Cosmology With Conformally Invariant Scalar Field \\(Physics Letters A 221 (1996) 359)}

\author{\sc{ Nivaldo A. Lemos}\\
\small{\it Departamento de F\'{\i}sica}\\
\small{\it Universidade Federal Fluminense}\\
\small{\it Av. Litor\^anea s/n, Boa Viagem }\\
\small{\it 24210-340, Niter\'oi, RJ - Brazil}\\
\small{e-mail: nivaldo@if.uff.br}}
\date{}
\begin{document}
\pagestyle{myheadings} 
\baselineskip 22pt
 
%\markboth{L.V.Belvedere  }{Supersymmetric Liouville Theory}
\maketitle
%\begin{abstract}

\vskip 2cm
The paper contains a harmless typographical mistake and, regrettably, an algebraic error with serious consequences. The prefactor in the last term of Eq.(2) should be 1/6 instead of 3, and the last term in Eq.(7) should include  $k$ as a factor. As a consequence, in the case $k=\Lambda =0$ considered in the paper Eq.(8) should read

$$p^2_R  - p^2_{\chi} = 0\,\,\, .$$
\\
\noindent In the second of Eqs.(9) the term $-4\chi^2$ within parentheses should be deleted, and the right-hand side of the last of Eqs.(9) should be zero. The classical equations of motion are solved by

$$R(t)=At\,\,\,\,\,\,\,\,\,\, , \,\,\,\,\,\,\,\,\,\, \chi (t)                       = \pm             At + B\,\,\, ,$$
\\
\noindent instead of       Eq.(10).
In Section 3 one should replace Eqs.(12) and (13) by

$$H=-p_R= \sqrt{p^2_{\chi} }\,\,\,\,\,\,\,\,\,\, , \,\,\,\,\,\,\,\,\,\, {\hat H}= \sqrt{{\hat p}^2_{\chi} }\,\,\, ,$$
\\
\noindent whereas Eqs.(14) and (15) should read

$$\varphi_p(\chi ) = e^{ip\chi}\,\,\,\,\,\,\,\,\,\, , \,\,\, \,\,\,\,\,\,\,{\hat H}\varphi_p(\chi ) =
\vert p\vert \varphi_p(\chi ) \,\,\,\,\,\,\,\,\,\, , \,\,\,\,\,\,\,\,\,\,
p\in R\!\!\!\!\! I\,\,\ .$$
\\

The reduced-phase-space discussion of singularities in the gauge $t=R$ is still valid, and Eq.(23) remains true {\it for small } $t$ if one just replaces $(n+1/2)/2$ by $\langle\varphi\vert {\hat\chi}^2\vert\varphi\rangle$ where $\varphi(\chi ) $ is a 
square-integrable wave function
at $t=0$.
The analysis based on the Wheeler-DeWitt equation, however, is now entirely different from the one given in the paper, which must be disregarded.
Although Eq.(25) is to be replaced by

$${\hat H}= -\sqrt{{\hat p}^2_{R} }\,\,\, ,$$
\\
\noindent the subsequent conclusions in Section 5 remain unchanged.

Thanks are due to Nelson Pinto-Neto, Jos\'e Ac\'acio de Barros e Marco A. Sagioro-Leal for pointing out the errors.

\end{document}